# Building Knowledge Graphs About Political Agents in the Age of Misinformation

Daniel Schwabe, Carlos Laufer, Antonio Busson
*Dept. of Informatics, PUC-Rio, R. M. de S. Vicente, 225, Rio de Janeiro, RJ 22453-900, Brazil*

**Abstract.** This paper presents the construction of a Knowledge Graph about relations between agents in a political system. It discusses the main modeling challenges, with emphasis on the issue of trust and provenance. Implementation decisions are also presented

Keywords: Political Systems, trust, linked data, provenance, influence networks.

## 1. Introduction

Although the term "Knowledge Graph" (KG) was introduced by Google in 2012[1], graph-based databases were available before this (e.g., Wordnet [28], DBPedia [22], Yago [38], CYC [23], NELL [7], and additional ones (e.g. ConceptNet [37]) continue to be created on a regular basis.

The majority of the largest published KGs are open-ended, in the sense that they include facts in practically any domain of knowledge. Consequently, for those having an underlying semantic model (e.g. DBPedia, Yago), their supporting ontologies are wide-ranging and are constantly being updated to accommodate new domains.

Another set of KGs are domain specific, and use specialized ontologies to describe their data ( e.g. [36] discusses several in the Life Sciences, [32] presents ontologies for the music domain).

KGs differ also on the way they are built (populated). A few are curated (e.g., CYC), others rely on crowdsourced information (e.g. Wikidata [39]), and most extract information from structured, semi-structured or textual information harvested from the Web.

The multiplicity of sources and various extraction approaches naturally raises the issue of data quality and confronts the user of the data in the KG with the issue of trusting, or not, the information obtained from the KG. For some types of information, for example in case of online reviews online and social media, this trust can have a direct effect on commercial success (e.g. [4]). This highlights the fact that data, ultimately, expresses a belief, opinion or point of view of some agent.

This paper presents an approach to building a KG in the domain of Political Agents, with a special emphasis on their different types of relations. This approach has been used in an initiative to build an open KG about Political Agents in Brazil in the form of Linked Data, named "Se Liga na Politica"[2] (SLNP). The data in this KG is obtained from several sources, in both automated and non-automated ways. Most of the automated extraction is made from official sources, such as the open data published by the House of Deputies and by the Senate. In addition to such sources, data may also be contributed by individuals, in crowdsourced fashion.

### *1.1. Political Systems*

Since very early times, men have organized themselves to form societies [34], naturally leading to the formation of Political Systems, defined as "formal and informal processes by which decisions are made concerning the use, production and distribution of resources in any given society". Societies include Agents – Persons and Organizations – that participate in the processes of its Political System. In doing so, they are driven, and constrained, by the various types of relations that exist among them.

[1]https://googleblog.blogspot.com/2012/05/introducing-knowledge graph-things-not.html

[2] The expression "Se liga" in Portuguese has a colloquial meaning of "be aware", "pay attention to", as well as "connect yourself". In Portuguese, it reads both as "pay attention to Politics" and "Connect yourself to Politics". A third (indirect) meaning is the reference to Linked Data.

With the advent of the Information Society and the Network Society [9] [13], accelerated by the widespread adoption of the Internet and the Web, information has become a vital resource, inextricably intertwined with the functioning of Political Systems.

Transparency, the quality that allows participants of the society to know what are the particular processes and agents that are being used in its functioning, is generally regarded as a means to enable checks and balances within Political Systems to prevent misuse by any of the parties involved [19] [26]. One of the forms to increase transparency within a Political System is to provide information about its participants and their relations, as a way to provide additional context when analyzing their actions, and, ultimately, making decisions.

In an ideal Linked Data World, a specific database about political agents would not really be necessary, as institutions responsible for each type of information would publish them in Linked Data form, creating a large KG. In practice, however, we are far from this – the majority of the published information about Political Systems is fragmented and incomplete. The focus of the SLNP project is to establish the *links* between the various "domains of knowledge" involved in describing Political Systems, taking care not to replicate all of the information published by each source (thereby, somehow, replacing it), focusing on characterizing the *relations* between agents and often omitting other properties that may be of interest of specific communities.

### *1.2. Trust*

Given the multiplicity of sources, and the nature of the subject matter, this KG is designed so that facts are seen as claims made by some agent, and therefore provenance information becomes a "first class citizen" of the domain. One of the main usages for this database is to provide context information for news stories, to allow readers to establish trust in the claimed facts based on their own criteria.

The issue of trust has been prevalent in the Internet since its popularization in the early 90s (see [16] for a survey), with a focus on the lower layers of the Internet Architecture, emphasizing authentication. More recently, with the advent of the Web and social networks, the cybersphere, and society as a whole, has become heavily influenced by information (and misinformation) that flows in news sites and social networks in the Internet. There are many studies carried out in several disciplines attempting to characterize and understand the spread of information in the cybersphere, and how this affects society (see [24] for an overview). A more prominent aspect has been the spread of "fake news", actually a term used to refer to several different misuses of information, as postulated by Wardle in [39]. This has also been the focus of much research and many initiatives (e.g. [10], [11], [3]).

The original vision for the Semantic Web included a "Trust" layer, although its emphasis was more on authentication and validation, and static trust measures for data. There have been many efforts in representing trust, including computational models - a general survey can be found in [31]; [5] presents an excellent earlier survey for the Semantic Web; and [35] surveys trust in social networks. In the Linked Data world, it is clear that facts in Semantic Web should be regarded as claims rather than hard facts (e.g., [6]), which naturally raises the issue of trust on those claims.

The remainder of this paper is organized as follows. Section 2 presents the Domain Model for relations beteen agents in Political Systems; Section 3 presents a model for the Trust Process supported by the SLNP KG, and details how provenance information is represented and used; Section 4 briefly discusses implementation aspects, and Section 5 draws some conclusions and points to ongoing and future work.

## 2. Domain Model

This section presents the POLARE ontologies that characterize the relations between agents in Political Systems. Before delving into details, some of the requirements for the ontology and the rationale for the design approach are discussed.

### *2.1. Methodological Approach*

#### *2.1.1. Ontologies vs vocabularies*

As a general rule, preference was given to using well-known ontologies, such as FOAF [3], ORG [4],

[3] http://xmlns.com/foaf/spec/

[4] https://www.w3.org/TR/vocab-org/

SKOS[5], Schema.org[6], etc… as controlled vocabularies to describe concepts in their respective domains. Precisely because these ontologies are very general, they allow many possible uses within other ontologies.

POLARE, in many situations, defines specific ways in which these vocabularies can be used for its purposes; whenever the intended use was incompatible with these ontologies, SLNP's own vocabulary was used. In addition, SLNP's vocabulary also includes terms to describe concepts not found in any of the better-known controlled vocabularies.

POLARE is meant to be used to characterize data in a Linked Data database. It is envisioned that this data may be used in my different ways, for various purposes. To allow such latitude, it was deliberately designed in a "lightweight" fashion, with few specific inference rules. It is understood that it is possible to extend it with a more "heavyweight" ontology by including inference rules to further constrain the possible interpretations, for use in specific situations.

One should also keep in mind that POLARE describes statements which are understood as claims being made by some agent (this is elaborated in section 3). Therefore, additional care must be taken when including inference rules, as they may be expressing restrictions according do some particular point of view, not necessarily accepted or agreed upon by all users.

The POLARE ontology includes several Datatype properties, but since they are not so relevant for characterizing relations, they will not be discussed in here.

#### *2.1.2. OWL vs SKOS*

OWL is a knowledge representation language, designed to formulate, exchange and reason with knowledge about a domain of interest. OWL can be reasoned with by computer programs either to verify the consistency of that knowledge or to make implicit knowledge explicit [18].

An alternative approach to represent knowledge is proposed by the SKOS ontology. A SKOS concept can be viewed as an idea or notion; a unit of thought. However, what constitutes a unit of thought is subjective, and this definition is meant to be suggestive, rather than restrictive [27].

The approach used for SLNP is based on a hybrid set of OWL vocabularies and SKOS Concept Schemes to describe concepts in its domain. OWL is used to define more formal structures where the inference rules can be used to make implicit knowledge. SKOS is used to define concepts that are mainly used for retrieval and navigation tasks, and for which there are many possible alternative schemes.

Several SKOS Concept Schemes have been identified that should complement the OWL classes defined in the POLARE OWL vocabularies. They are used as classifications for specific classes. The rationale for choosing to use SKOS as opposed to OWL was based on the generality vs specificity of the concept involved – whenever the concept could be represented in many different ways depending on the particular Political System, SKOS was preferred. For example, the "classification" of an Organization can be made in many different taxonomies, often non-mutually exclusive – for instance, according to fiscal status, legal status, type of ownership/control, etc… Such uses will be highlighted throughout the description of POLARE. Whenever possible, preference was given to utilizing standard vocabularies.

### *2.2. Domain Models*

#### *2.2.1. People and Organizations*

The central concepts in POLARE are Persons and Organizations, as they are the Political Agents within a Political System. Given the goal to characterize the various kinds of relations between them, direct relations between Persons was the first focus, and then relations between Persons and Organizations were examined, as they establish indirect relations between Persons. The FOAF vocabulary was chosen to describe Persons, and the ORG vocabulary to describe Organizations, adding relations in the POLARE ontology as needed[7].

The first kind of relations between persons are direct family relations., which are modeled in POLARE as Direct Relationships, shown in **Error! Reference source not found.**. Rather than simply using an *owl:ObjectProperty*, they are modeled via reification, due to the need to qualify this relation with temporal information. The *directRelProp* property allows specifying what is the family relation; its value, rather than being an *rdf:Property*, is a *skos:Concept*, whose value will be taken from a suitable *Skos:ConceptScheme*. This allows inclusion of certain relations that may not be "formally" accepted as a family relation but may be of interest for some types of analyses, e.g., "co-habitates".

[5] http://www.w3.org/TR/skos-reference
[6] http://schema.org/docs/full.html

[7] For readability purposes, we do not add a prefix to terms of POLARE itself (e.g., pol:hasPost). Similarly, when it is clear which ontology a term is from (e.g. foaf:Person), we omit the prefix.

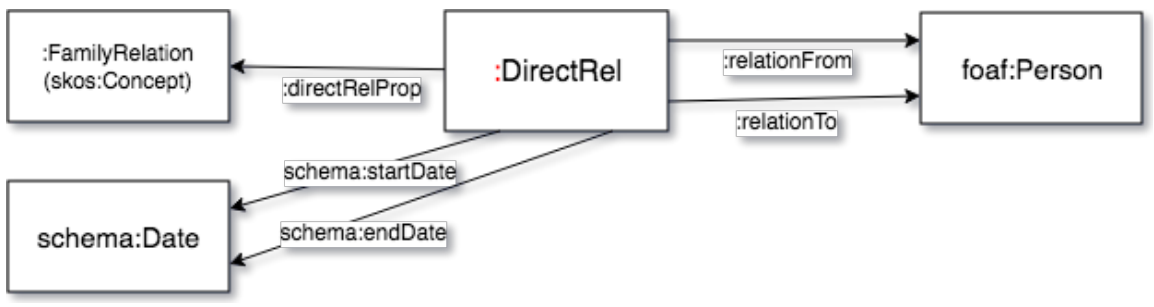

Fig. 1 – Direct relationships between Persons in POLARE

Direct relations between Organizations are already contemplated in the ORG vocabulary (e.g., *org:hasSubOrganization*, *org:HasUnit*). In the future, relations between Organizations may be extended in an analogous way as the *DirectRel* approach, to represent more nuanced relations such as shareholding control.

Persons are related to Organizations chiefly through occupying positions within them, modeled using the *org:Post* class[8]. The main terms of the POLARE ontology come from W3C's ORG Ontology, shown in Fig. 2. It is important to notice the reification of the relation *occupies* between *Person* and *Post* via the *Membership* class, which is fundamental to allow representing properties of this relation, as will be discussed later.

The property *org:role* is used to identify an *org:Post* occupied by a person in an *org:Organization*.; its value is an *org:Role* that is a *skos:Concept* defned in a specific SKOS Concept Scheme.

The *hasPost* relation linking *org:Membership* with *org:Pos*t has been added because of the need to represent that a Post might be occupied by different persons in different periods of time.

Both *Post* and *Membership* may have start and end dates associated to them. The dates associated to a *Post* refer to a time period when the *Post* exists, for example, for a post in the House of Representatives, it corresponds to a Legislature, which defines the mandate of the elected person. The *Membership* dates refer to the period in which the *Person* actually occupies the *Post*, since it is possible for a *Person* to temporarily leave the *Post* for a period of time (leave of absence), e.g. within a legislative mandate.

An important characteristic of most public organizations that make up a Government is that they have fixed number of Posts. For example, the House of Representatives has a fixed number of seats (*Posts*), and a particular seat can be occupied by at most one person in any given moment. For this reason, although the ORG ontology allows direct *org:memberOf* relations between Agents and Organizations, it was decided not to use this relation, requiring always an *org:Post* to exist to mediate the relation between *Persons* and *Organizations*.

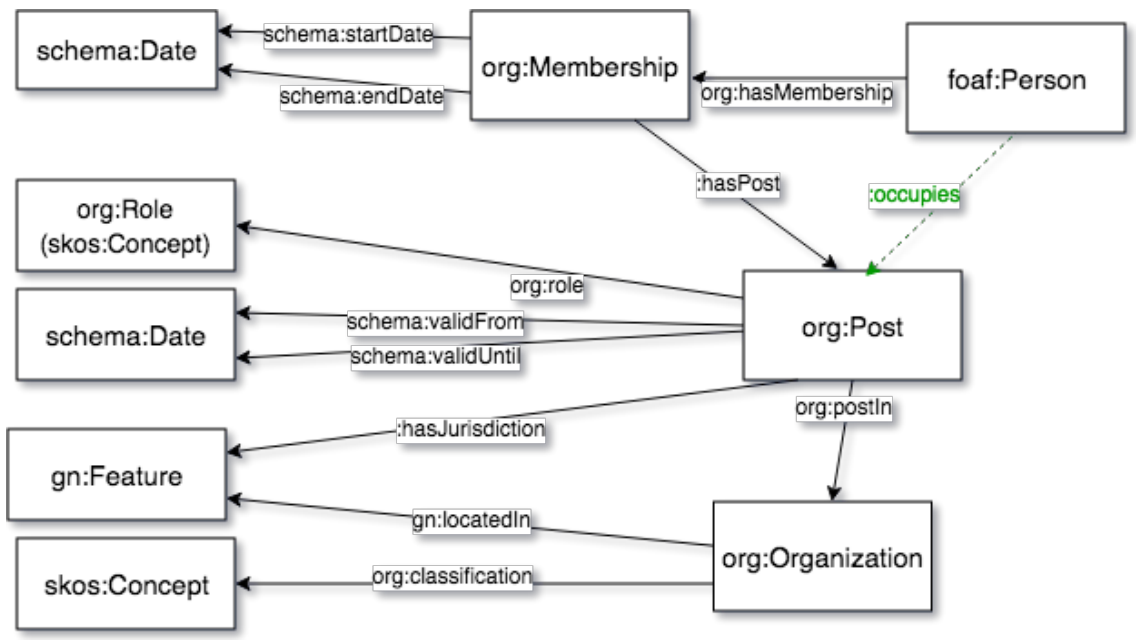

Fig. 2- Persons and Organizations in the POLARE ontology. Inferred relations are shown as dotted lines.

In Political Systems, and also for many Organizations, it is important to know if some Post was filled through a *Referral*, i.e., that some *foaf:Agent* indicated (nominated) a *foaf:Person* to occupy an *org:Post*. POLARE can represent which *Agent* has referred some *Person* to occupy a *Post* in an *Organization*, shown in Fig. 3. The *refers* property is reified using a *Referral* class, to allow including properties related to the Referral itself.

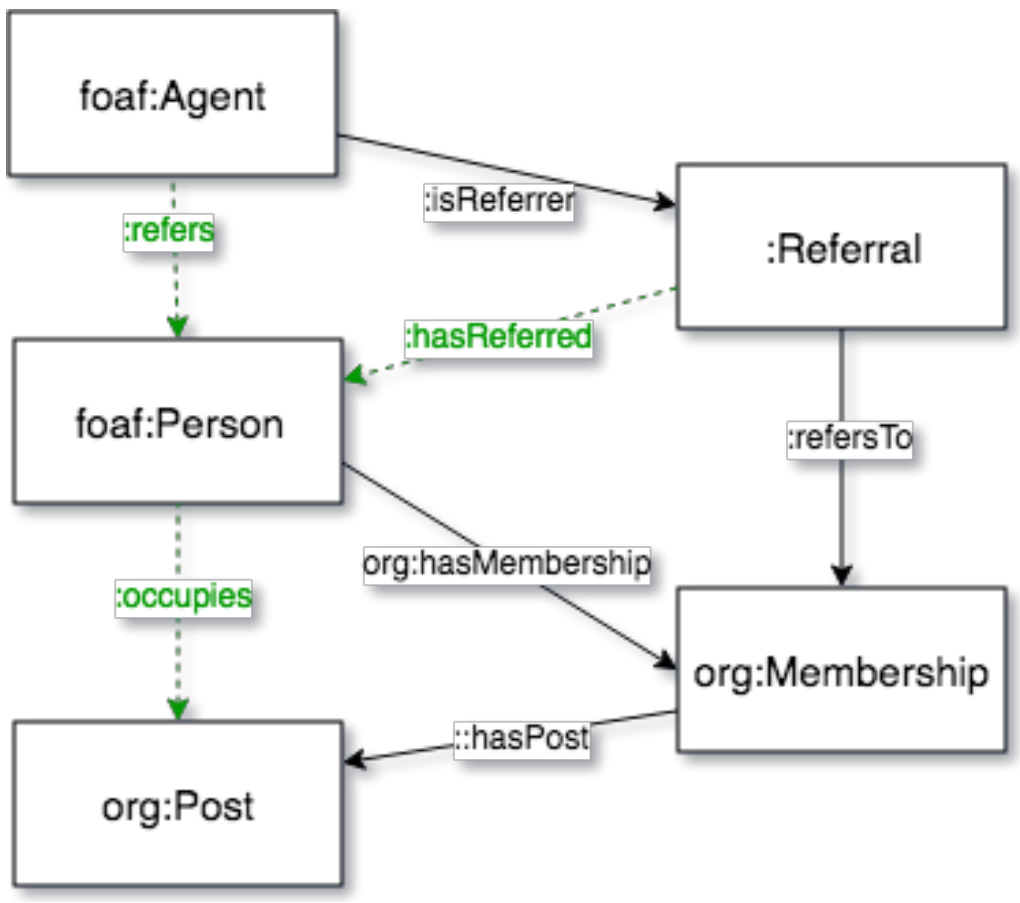

Fig. 3 – Modeling Referrals (nominations) in POLARE

Several indirect relationships of interest between Persons can be captured simply by the fact that these Persons occupy Posts in the same Organization. For example, the fact that two congressmen belong to the same party (which is an Organization), or that they were once colleagues in a company in the private sector or in a department in the executive branch.

[8] Notice that "Post" here refers to a "Position", not to a blog post.

*2.2.2. Legislative aspects*

One of main activities in a Political System is the elaboration and enactment of legislation that establish and regulate many of its processes. Fig. 4 shows a diagram of the part of POLARE showing relations involved in these activities.

Legislators (*foaf:Persons*) make (*dc:creator*) *Propositions* (known by several terms in particular Political Systems, e.g., "Bill" in the US and Uk, "Projeto de Lei" in Brazil, "Projet/Proposition de Loi" in France) that undergo a voting process to become a *Law*. The voting process takes place in *Sessions* (which are *schema:Event*s) comprised of a series of *VoteEvent*s. Each *VoteEvent* requires legislators (*Voter*) to issue a *Vote* regarding some *Disposition* (e.g., substitution, amendment, approval, etc…) relative to the *Proposition* in question. The actual vote (e.g., "yes", "no", "abstain", etc…) is modeled as a *skos:Concept* that is the value of the *vote* property for instances of class *Vote*.

The class *Voter* stands for the role of the *Person* in voting and is necessary because it is important to preserve the relation with the Party (which is an *Organization*) to which the legislator belongs. Since in some Political Systems legislators can change their party affiliations during the course of their mandate (as is the case in Brazil), it is important to record the particular affiliation in effect at the time of the vote. It is true that this information could be inferred by retrieving the set of instances of *Membership* between the *Person* and Political Parties (as instances of *Organizations*), checking each for its associated time interval (*startDate* and *endDate*), and determining the one that was in effect at the *startDate* of the *VoteEvent.* Nevertheless, it was decided to include this information directly in the recorded data, since it is directly provided by the datasources used in the project and is also commonly provided by datasources reporting legislative activity worldwide.

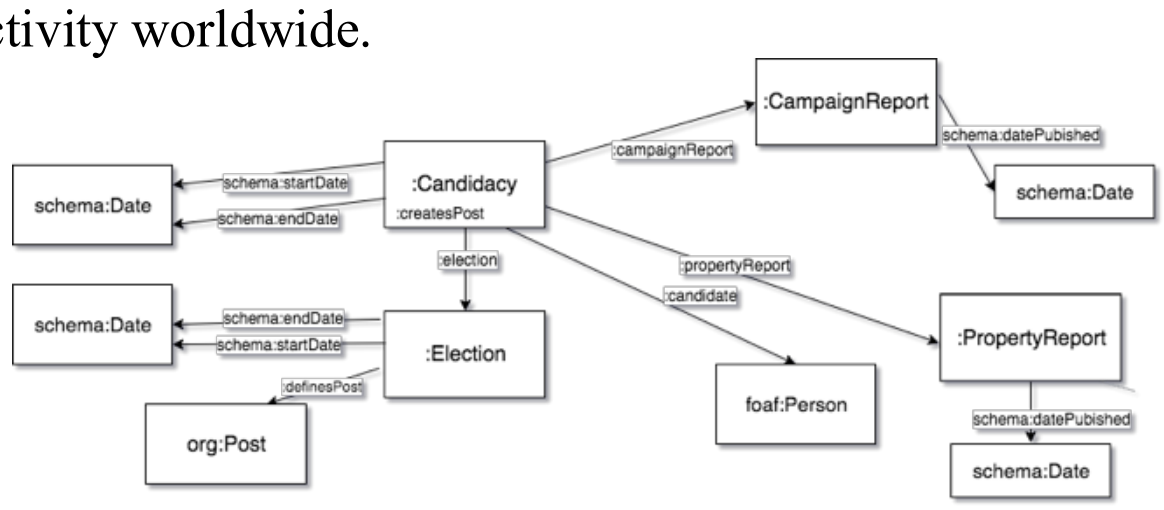

Fig. 4 –Relations derived from legislative activity in POLARE

For many Political Systems, certain *VoteEvent*s may have a voting Recommendation issued by some group of voters (*foaf:Group*), for example, a caucus or a political party.

*2.2.3. Electoral Process*

A central aspect of a Political System is the way public officials are chosen, and involvement in this process is another indirect way in which relations between Political Agents are established. Fig. 5 shows a portion of POLARE modelling the electoral process.

An *Election* defines a number of *org:Post*s (e.g. seats in the Senate) for which Persons can run for. In other words, a *Candidacy* models that a *foaf:Person* is a *candidate* in an *Election* for an *org:Post*. Notice that although it is expressed in terms of an election, this relation pattern could be applied to many other selection processes.

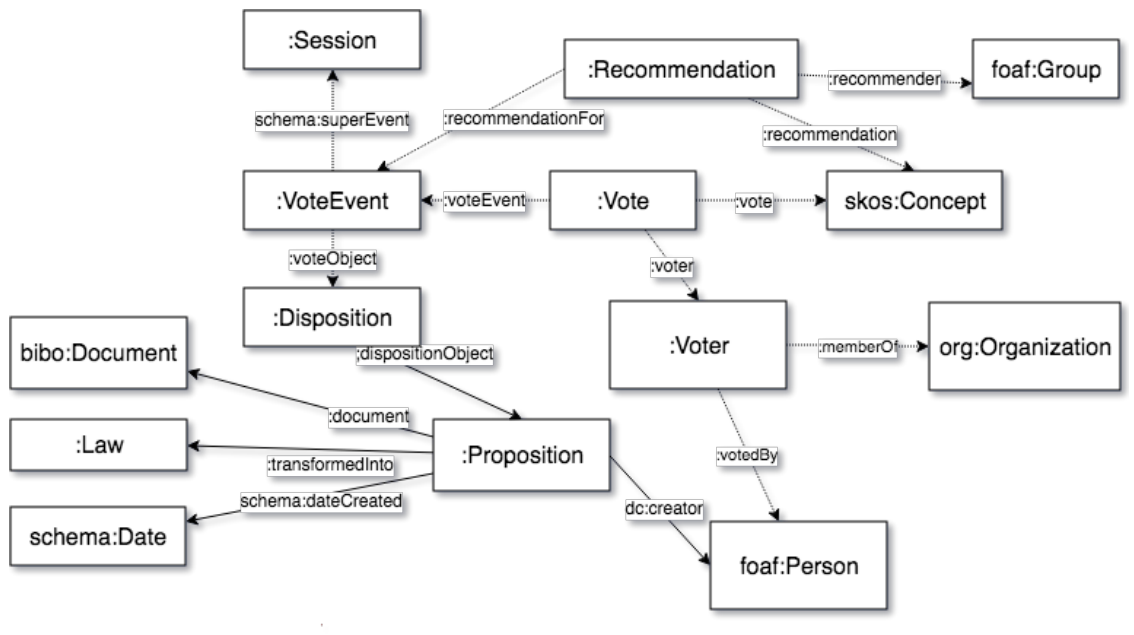

Fig. 5 Electoral activity in POLARE

For many Political Systems, transparency requirements include filing a *CampaignReport* to detail the finances (incomes and expenses) incurred during the electoral campaign. Another requirement is the filing of a *PropertyReport* detailing the assets owned by the candidate at the time of the candidacy. It is expected that this report can be compared to a similar one issued when the person leaves the office, enabling verification of possible irregularities.

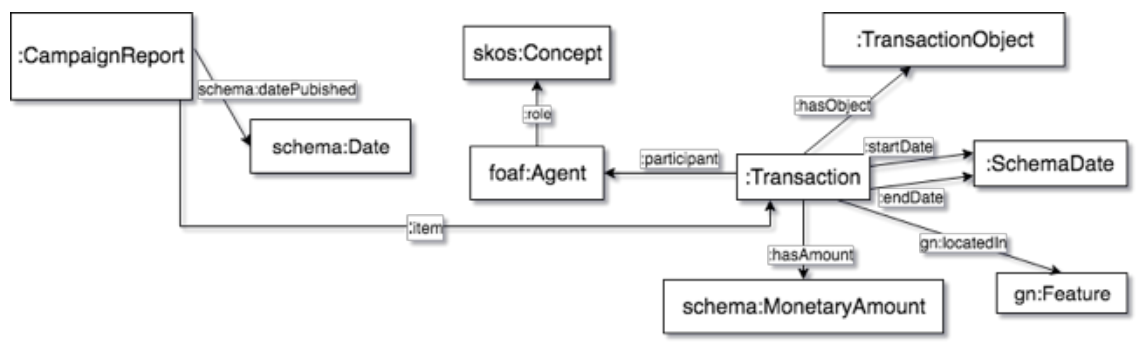

Fig. 6 Transactions in POLARE

The campaign financial report is a list of *Transaction*s. Each *Transaction* involves a number of *foaf:Agents* each assuming a *role* (e.g., seller, buyer, guarantor, etc…) which is *skos:Concept* chosen from a suitable *skos:ConceptScheme*. The transaction involves an object (*TransactionObject*, which is either a *schema:Product* or *schema:Service*) that is rendered possibly in exchange for a certain amount.

It should be noted that although this model for transactions is presented here within the context of a campaign financial report, it may be used to record any transactions involving Political Agents, not necessarily during an electoral campaign. Thus, two or more Political Agents are (indirectly) related if they participated in the same *Transaction*.

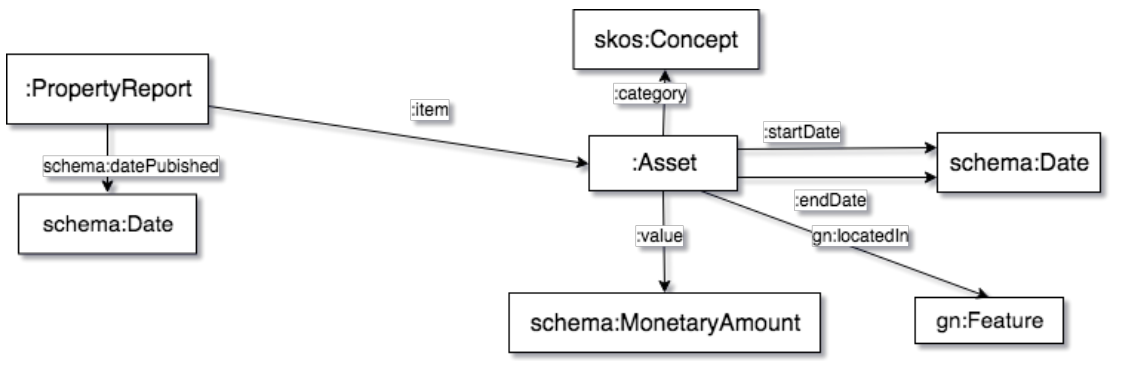

Fig. 7 Asset list in POLARE

Fig. 7 shows the POLARE fragment representing a *PropertyReport,* which is simply a list of *Assets* owned by the candidate in an *Election*. Note that, this list will only characterize relations between Political Agents under the presence of some *Transaction* that establishes how that *Asset* came into possession of it owner. As such, an *Asset* is related to *TransactionObject.*

### 2.2.4. *Legal actions*

Another important aspect of Political Systems is the way it handles violations of its established norms. When such a situation occurs, it is dealt with through legal actions.

Fig. 8 shows the fragment of POLARE representing the relations involved in a *LegalCase*.

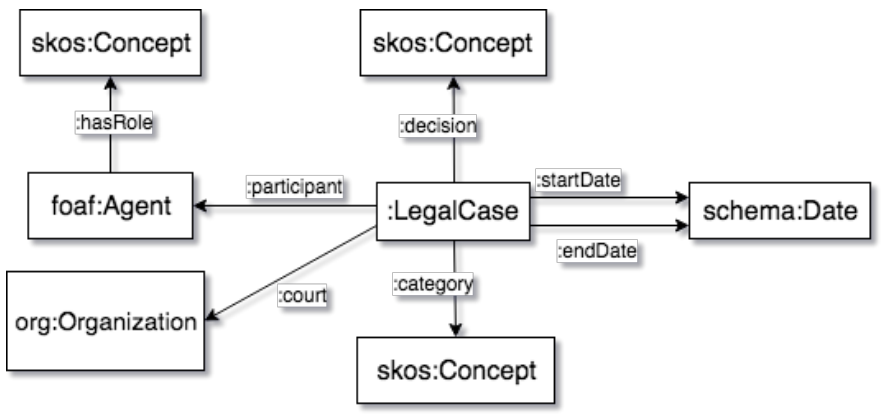

Fig. 8 – POLARE representation of legal cases

Each participant is a *foaf:Agent* that fulfills a role (*hasRole*) in the case, such as plaintiff, defendant, judge, attorney, etc…Once more, such roles are modeled as *skos:Concept*s in a *skos:ConceptScheme*, which must exist for the various Political Systems. Thus, two Political Agents are (indirectly) related if they participated, in any role, in the same *LegalCase.*

### 2.3. *Using SHACL*

The POLARE ontology provides a set of vocabularies to represent the relations among Political Agents by composing concepts from existing vocabularies, and adding specific classes and properties as needed. In order to characterize the particular composition patterns intended to be instantiated in the knowledge graph, the same information in the preceding diagrams is expressed using the Shapes Constrain Language recommended by the W3C[9].

Fig. 9 shows an example of the expression that represents the diagram in Fig. 2.

```
:PersonShape a sh:NodeShape ;
    sh:targetClass foaf:Person ;
    sh:ignoredProperties ( rdf:type ) ;
    sh:property [sh:path foaf:name ;
        sh:datatype xsd:string ;
        sh:minCount 1 ;] ;
    sh:property [sh:path org:hasMembership ;
sh:node org:Membership ;
    ] .
:MembershipShape a sh:NodeShape ;
    sh:targetClass org:Membership ;
    sh:closed "true"^^xsd:boolean ;
    sh:ignoredProperties ( rdf:type ) ;
    sh:property [sh:path schema:startDate ;
        sh:datatype schema:Date ;
        sh:minCount 1 ;sh:maxCount 1 ;] ;
    sh:property [ sh:path schema:endDate ;
        sh:datatype schema:Date ;
        sh:maxCount 1 ;] ;
    sh:property [sh:path agent:hasPost ;
        sh:node org:Post ;
    ] .
:PostShape a sh:NodeShape ;
    sh:targetClass org:Post ;
    sh:closed "true"^^xsd:boolean ;
    sh:ignoredProperties ( rdf:type ) ;
    sh:property [sh:path schema:startDate ;
        sh:datatype schema:Date ;
        sh:minCount 1 ;sh:maxCount 1 ;] ;
    sh:property [ sh:path schema:endDate ;
        sh:datatype schema:Date ;
        sh:maxCount 1 ;] ;
    sh:property [sh:path org:postIn ;
        sh:minCount 1 ; sh:maxCount 1 ;
        sh:node org:Organization ;] ;
    sh:property [ sh:path org:role ;
        sh:minCount 1 ;sh:maxCount 1 ;
        sh:node skos:Concept ;
    ] .
:OrganizationShape a sh:NodeShape ;
    sh:targetClass org:Organization ;
    sh:ignoredProperties ( rdf:type ) ;
    sh:property [sh:path skos:prefLabel ;
        sh:datatype xsd:string ;
        sh:minCount 1 ; sh:maxCount 1 ;
   ] .
:SkosConceptShape a sh:NodeShape ;
    sh:targetClass skos:Concept ;
    sh:ignoredProperties ( rdf:type ) ;
    sh:property [sh:path skos:prefLabel ;
        sh:datatype xsd:string ;
        sh:minCount 1 ; sh:maxCount 1 ;] ;
```

[9] https://www.w3.org/TR/shacl/

```
    sh:property [sh:path skos:broader ;
        sh:node skos:Concept ;] ;
    sh:property [
        sh:path agent:inOrgRoleScheme ;
        sh:node skos:ConceptSheme ;
        sh:minCount 1 ;] ;
    sh:property [
        sh:path skos:topConceptOf ;
        sh:node skos:ConceptSheme ;
] .
:SkosConceptSchemeShape a sh:NodeShape ;
    sh:targetClass skos:ConceptScheme ;
    sh:ignoredProperties ( rdf:type ) ;
    sh:property [sh:path skos:prefLabel ;
        sh:datatype xsd:string ;
        sh:minCount 1 ; sh:maxCount 1 ;] ;
    sh:property [
        sh:path skos:hasTopConcept ;
          sh:node skos:ConceptSheme ;] .
```

Fig. 9 – ShaCL expression for Person and Organizations as depicted in Fig. 2

## 3. Trust

Whereas there are many definitions of trust (e.g. [25]), the approach used is based on the work of Gerck [14] and Castelfranchi et al. [7], taking the view that trust is "knowledge-based reliance on received information", that is, an agent decides to trust (or not) based solely on her/his knowledge, and the decision to trust implies the decision to rely on the truth of received or known information to *perform some action*.

Castelfranchi et al. define trust in the context of multi-agent systems, where agents have goals, asserting that trust is "a mental state, a complex attitude of an agent x towards another agent y about the behavior/action relevant for the result (goal) g. This attitude leads the agent x to the decision of relying on y having the behavior/action, in order to achieve the goal g".

Gerck presents a definition of trust as "what an observer knows about an entity and can rely upon to a qualified extent". The two definitions are closely parallel: the observer is the agent who trusts; the entity is the trusted agent; the qualified extent is the behavior/action. Both associate trust with reliance. However, the former definition mentions explicitly the goal-oriented nature of trust.

From both definitions, it can be observed that trust implies reliance: when an agent trusts something, it relies on its truth to achieve some goal without further analysis – even if it is running the risk of taking an inappropriate or even damaging action if the object of trust is false.

### *3.1. A Framework for the Trust Process*

Given the considerations above, we present a model for the trust process that underlies the use or consumption of data/information on the web, represented diagrammatically in Fig. 10.[10]

We focus here on the cases where an *Agent* needs to act, i.e., do some computation, make a decision, or take some *Action*. The agent must act based on some *Data/Information* items which, clearly, it must trust – the *Trusted Data*. The *Data/Information* items to be used by the agent may come from several sources, and it is not always clear (to the *Agent*) what is the quality of this *Data*, or the trustworthiness of the *Information* it contains. Therefore, the *Agent* must apply a *Trust Process* to filter the incoming *Data/Information* items and extract the *Trusted Data* items to be used by the *Action*. From this point of view, an item is considered as the smallest indivisible element that can be used in the *Trust Process* and may have an internal structure when used by the *Action*.

This *Trust Process* can be based on a multitude of different signals, some of which we have singled out in the diagram in Fig. 10, namely, the *Metadata* which describes various properties of the *Data/Information* items, and the *Context* in which the *Action* will take place. The criteria used in the *Trust Process* are expressed by *Policies* determined by the *Agent*. Notice that in this framework, the *Context* contains any arbitrary information items used by the *Policies*, in addition to *Metadata* and the *Data/Information* items themselves.

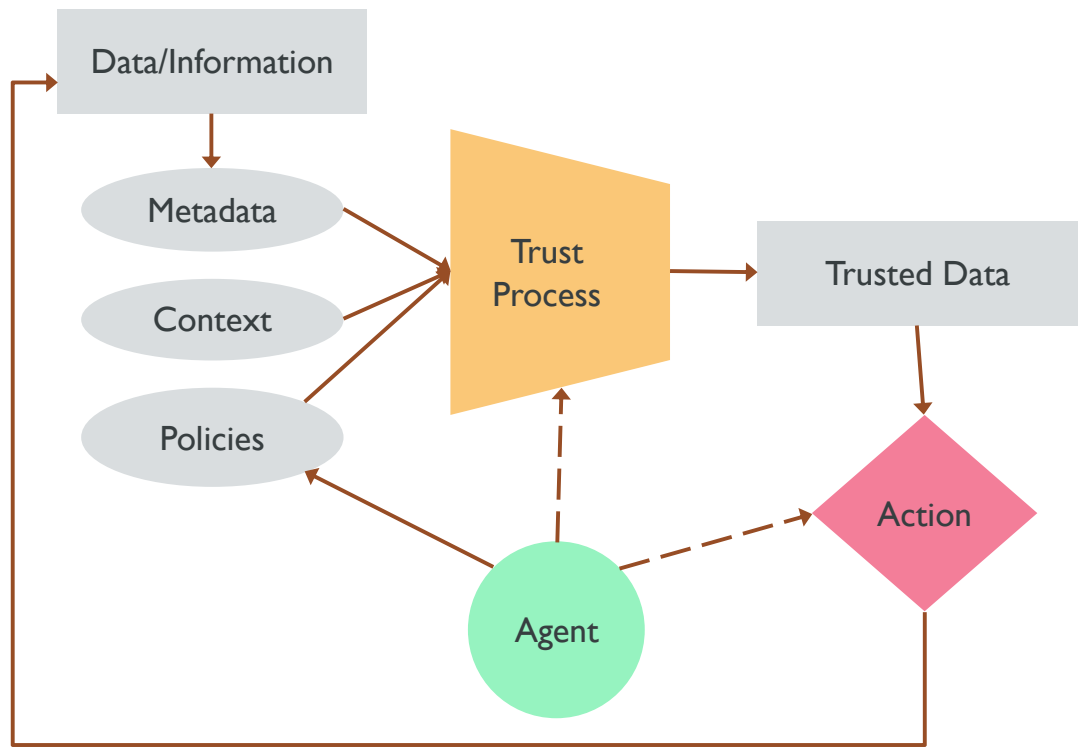

Fig. 10 – A Framework of the Trust Process. Continuous lines represent "consume/produce"; dashed lines represent "executes".

Many, if not most, of the trust models discussed in the surveys mentioned in Section 1 [5],[35] can be regarded as providing models or representations for one or more of the elements of this trust framework, such as different representations for the metadata, or specification language for the policies, or particular types of context information that can be added to the data. Regardless of these models, it should be clear that, as far as the *Action* is concerned, the whole process is *binary*: An incoming *Data/Information* item is either accepted and added to the *Trusted Data*, or it is not – there are no “half filters”. In other words, when time comes to *Act*, either the *Agent* uses that *Data/Information* item, or it does not, it can’t “use it partially” – an element is considered indivisible from the point of view of the *Trust Process* [2].

[10] This section has been extracted with some adaptation from [20]

Thus, the actual trusting process (“Model of Trust”) should not be confused with models which may attribute non-discrete or continuous values to the trustworthiness assigned to data/information items that are used to determine if the incoming data/information items should be filtered or not (“Trust Models”).

Another point to notice is that, in this framework, only the *Agent* determines the policies it wishes to apply to incoming *Data/Information*. The metadata associated with the incoming *Data/Information* may contain *Data/Information* about the publisher or provider of a *Data/Information* item, and the *Policies* may take this into account in the filtering process.

In contrast, privacy issues would require adding a similar set of metadata and policies to the *published Data/Information* that will be consumed by the *Agent*, which would act as an additional filter, applied before the *Data/Information* is made available to the *Agent*. The privacy *Policies* of a Publisher may use information about the (requesting) *Agent*, the *Action* and the *Context*.

### *3.2. PROVHeart - a Claim Provenance ontology pattern*

As already discussed, a critical information to establish trust is data related to the provenance of that information, since acceptance of the truth of a claim relies on the trust one has on the Agent making the claim. For the Political Systems domain, the focus is on the relationships between agents, which are modeled in POLARE using the *Membership*, *Direct Relation* and *Referral* classes.

The PROV-O Ontology[11] provides a vocabulary to represent provenance information, which can be used in a variety of ways to represent diverse aspects of provenance. For our purposes, especially because we allow individual users to enter information (including provenance) in the database, it is important to ensure that the provenance information is recorded in a uniform way, providing enough information to allow the trust process. To achieve this, the notion of a pattern is used, similar in spirit to the PROV template proposed in [29]. The advantage of using PROV Patterns is to extend the ontology with a well-formed structure using the PROV vocabulary, allowing the construction of queries that use PROV data mixed with the data of the main ontology.

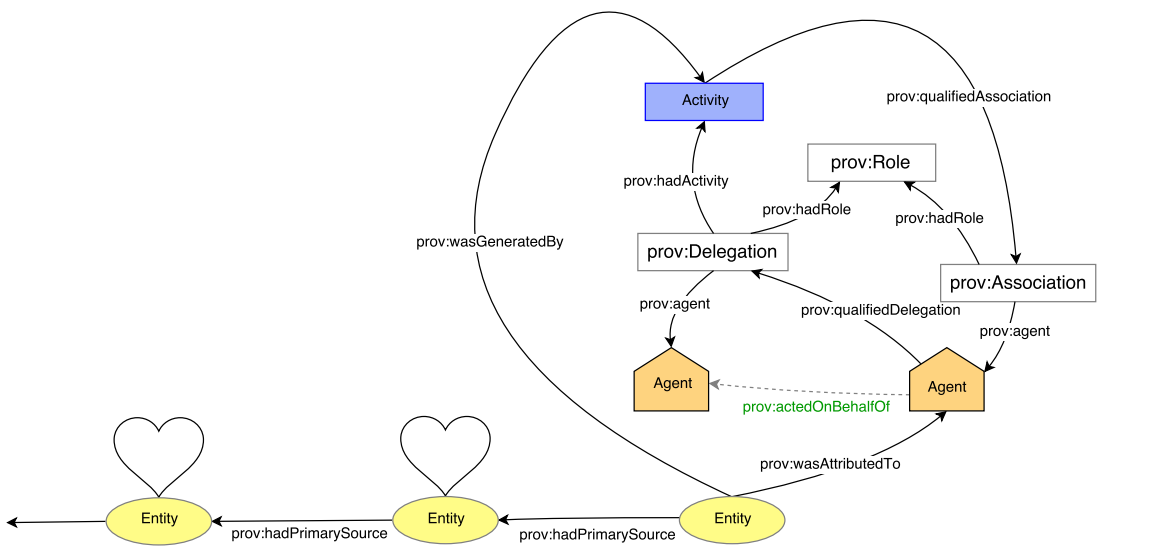

Fig.11 – The PROVHeart Provenance Pattern

We define a provenance pattern called PROVHeart, shown in Fig.11. PROVHeart is a simple pattern that models the provenance of an Entity that was generated by an Activity attributed to an Agent with a Role in that Activity, using another Entity as a source, and acting on behalf of another Agent. Provenance chains can be built using the *prov:hadPrimarySource* relation between Entities that serve as the “justification” or “basis” for each claim, forming the basis for the trust chain discussed earlier. Notice also that “*prov:actedOnBehalfOf*” is represented as a property chain property, because this relation has to be linked to the Activity that produced the particular Entity the provenance refers to. In other words, “*prov:actedOnBehalfOf*” is seen as applying to specific activities, and not as a “blanket statement” that *always* holds for all Activities carried out by an Agent, relative to another Agent.

A provenance chain can be established using the *prov:hasPrimarySource* property. The “heart” symbol represents an instance of this pattern.

[11] https://www.w3.org/TR/prov-o/

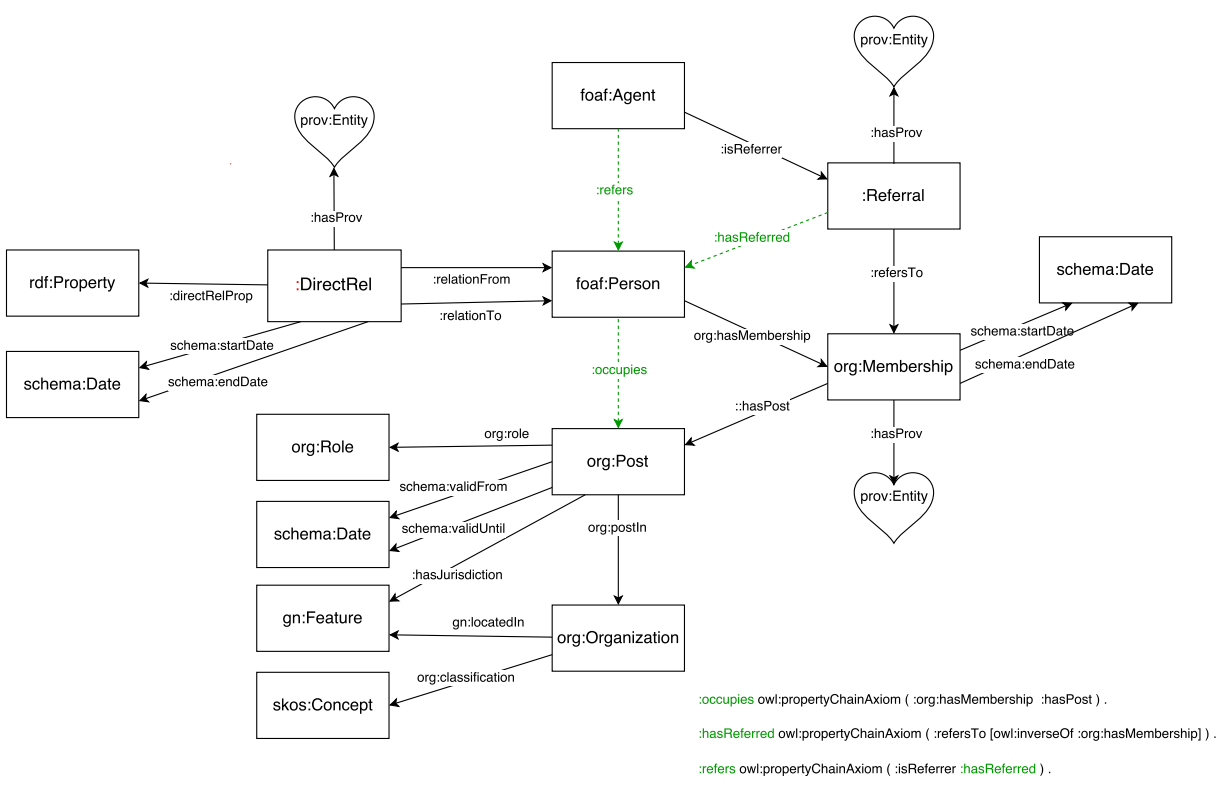

Fig. 12 – The complete POLARE ontology including provenance information.

Fig. 12 shows the complete POLARE ontology, where the heart symbol represents the PROVHeart pattern. One can see that claims involving any of the main relations (Membership, Referral and DirectRel) have provenance information attached to them.

As an example of PROVHeart use, consider the case where someone wants to know all Persons, and respective Posts, that were referred by a Person called “Jader Barbalho”. Furthermore, these Referrals should have provenances stating that the sources used (evidence) were published by a Person called “Andréia Sadi” or are from a Magazine called “Época”, and, also, should be published to the database by a Person called “Daniel” or “Laufer”.

In the implementation used, claim representation is based on the nanopublication model [17][12], where a claim, represented by a set of statements, is recorded as a group of three named graphs:

– assertion graph - containing the set of statements that compose the claim, using POLARE ontology;
– publication info - containing statements that give provenance information about the claim publication activity, using the provHeart pattern;
– provenance - containing statements that give provenance information about the claim itself, using the PROVHeart pattern.

In implementation, a named graph called *claims* is used to store the nanopublications related to the set of all claims published in the database.

The example above can be queried in the database using the following query:

```
  SELECT DISTINCT ?referrerName ?referred-
Name ?roleName ?startdate ?citation
 WHERE {
   GRAPH graph:claims {
     ?pub np:has Assertion
 ?claim ;np:hasProvenance ?prov ;np:hasPublication-
Info ?pubinfo . }
    GRAPH ?claim {
      ?referral a polare:Referral ;polare:hasRe-
ferred ?referred ;schema:startDate ?startdate .
      ?referrer polare:isReferrer ?refer-
ral ;foaf:name ?referrerName .
      ?referred polare:occu-
pies ?post ; foaf:name ?refferedName .  ?post
org:role ?role .
      ?role skos:prefLabel ?roleName .
      FILTER (?referrer = ex:JaderBarbalho) . }
    GRAPH ?prov {
      ?claim prov:hadPrimarySource ?source ;
prov:wasAttributedTo ?agent ;
                  prov:wasGeneratedBy ?activity .
      ?source dcterms:bibliographicCitation ?cita-
tion .
      ?delegation prov:hadActivity ?activ-
ity ;prov:agent ?org .
      FILTER (?agent = ex:AndreiaSadi || ?org =
ex:EpocaMagazine) . }
    GRAPH ?pubinfo {
      ?pub prov:wasAttributedTo ?agpub .
      FILTER (?agpub = ex:Daniel || ?agpub =
ex:Laufer) . }
```

In terms of the Trust Framework, the claims discussed here allow describing Data/Information items and Provenance information about them.

It should be noted that such provenance information will naturally be used in particular trust policies, seen as rules or constrains to filter incoming data graphs, before they are added to the trusted information database. Examples of such policies can be found in [2].

PROVHeart can be adapted to represent provenance information for bulk ingestions of data, by making use of the *prov:Plan* concept, which can be used to described the process followed.

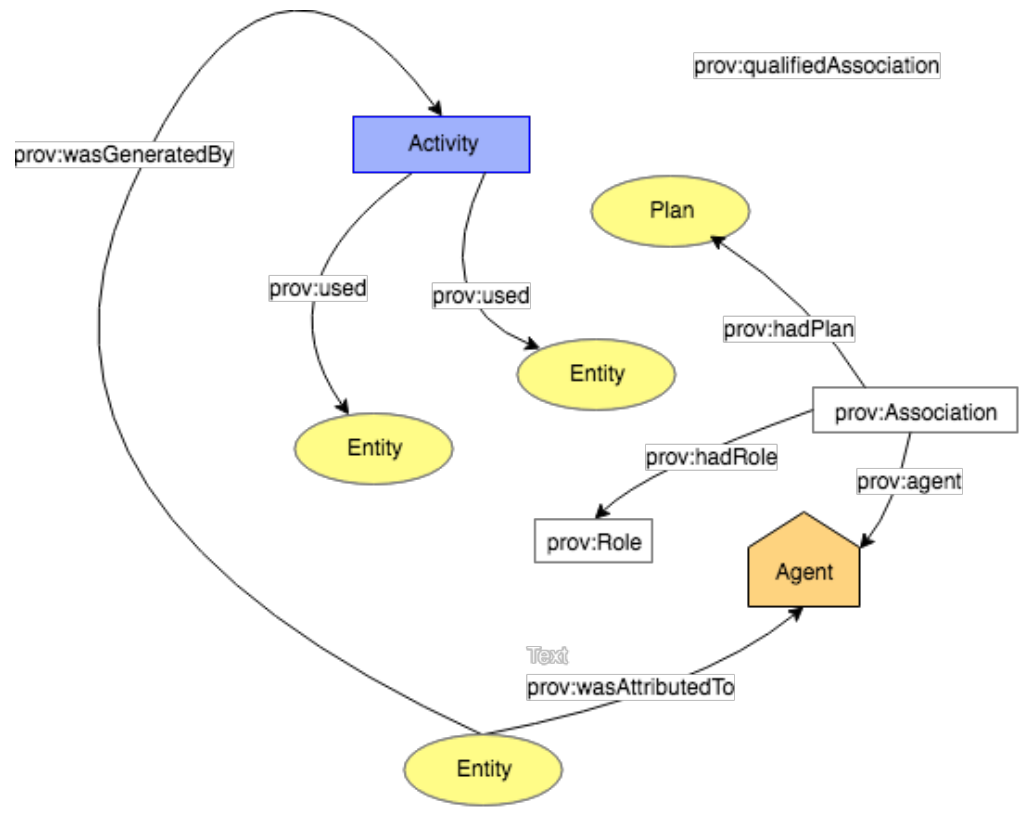

Fig. 13 – PROV Heart adaptation to represent a bulk import process.

[12] http://www.nanopub.org

An example of instantiating this pattern for bulk importing voting data from the data dump available for the House of Representatives information is given in Fig. 13.

```
:voting a prov:Entity;
   prov:wasAttributedTo :laufer;
   prov:wasGeneratedBy :importing2014.

:importing2014 a prov:Activity;
   prov:qualifiedAssociation [
      a prov:Association;
      prov:agent   :laufer;
      prov:hadRole :researcher;
      prov:hadPlan :importVotePlan;
   ];
   prov:used :voting2014;
   prov:used :karmaModel.

:laufer a prov:Agent.

:importVotePlan a prov:Plan, prov:Entity;
   rdfs:comment "Describes the ETL to generate the
voting rdf triples"@en;
   dcterms:URI res:importVotePlan.pdf.

:researcher a prov:Role .
:voting2014 a prov:Entity;
   rdfs:comment "CSV file containing the 2014 vot-
ing"@en;
   dcterms:URI res:voting2014.csv.

:votingKarmaModel a prov:Entity;
   rdfs:comment "Karma file to transform a csv file
with voting information to rdf triples"@en;
   dcterms:URI res:votingKarmaModel.ttl.
```

Fig. 14 – Provenance of bulk import of voting data using PROV Heart variant in Fig. 13.

In this example, a *prov:Plan*, *importVotePlan* defines the plan used for the process by providing a link to a PDF file with the description (meant for human consumption). It uses a software script represented by *votingKarmaModel*, which in turn refers to a formal description contained in a .ttl file. Both the plan and the model have provenance information of their own, not shown.

### *3.3. POLARE in action*

This section presents an example of how the database can be used to better understand a news story, discussing the options made available to the reader to establish her/his trust on the information contained in the story.

Consider the blog post[13] made by reporter Andréia Sadi, who works for the GloboNews news channel, about an investigation carried out by the Federal Police in the office of Congresswoman Simone Morgado, wife of Senator Jader Barbalho. This investigation is related to embezzlement accusations against an aide working in the Congressswoman's office. Fig. 15 contains a simplified diagram with the main facts regarding the Persons, Posts, and Organizations cited in the blog post. In this diagram, we use a simplified view of provenance, showing only a few provenance facts using the "source" link.

The information in the SNLP database contains the following information about the persons involved in the blog post.

Jader Barbalho is a Brazilian lawyer, politician and businessman. In his long political career, he was a Federal Deputy of Pará, Governor of Pará, Minister of Social Security of Brazil and Minister for Agrarian Development of Brasil. He occupies the post of Senator of Pará in the current legislature. He was married to Simone Morgado, until the end of 2016, according to his son, Helder Barbalho.

Helder Barbalho is a Brazilian politician and administrator. He was, during 8 years, the Mayor of Ananindeua, a city located in the state of Pará. He was the former Minister of Fisheries and Aquaculture of Brazil and is currently the Minister of National Integration.

Simone Morgado is a Brazilian economist and politician. She was Mayor of Bragança, a city located in the state of Pará and is currently a Congresswoman representing the state of Pará.

Soane Castro currently occupies the post of aide in the office of Simone Morgado, and held the post of Superintendent in the Ministry of Fisheries and Aquaculture of Brazil, during the period when Helder Barbalho was its Minister.

The shaded area indicates facts (actually, claims) that are not directly cited in the story, but are part of the database. In particular, it explicates that there are several connections linking Senator Jader Barbalho to the aide who is the target of the investigation – she was appointed for the position related to the embezzlement accusations by the Senator's son, Helder Barbalho, who was the Minister in charge of that agency at that time. This post (position) is in the home state of all agents involved, Pará.

Consider now the kinship information about Simone Morgado and Jader Barbalho. The blog post itself states that they are married, but also adds that

[13] http://g1.globo.com/politica/blog/andreia-sadi/post/pf-realiza-operacao-na-camara-dos-deputados-mas-alvo-nao-e-localizado.html. (In Portuguese)

Helder Barbalho claims they have not been married since November 2016. The fact that the reporter did not directly refer to Ms. Morgado as "Senator Jader Barbalho's ex-wife" seems to indicate she chose not to fully trust the information that they are divorced, and instead refers the reader to a secondary source citing the Senator's son claim about this fact. However, she did not give any provenance to Helder Barbalho's claim – i.e., how did she become aware of the claim. Fig. 16 shows the POLARE instance for the kinship relation claimed in the blog post, including reference to PROVHeart instances, Prov_2 and Prov_3.

Assuming that agent Laufer was responsible for entering the facts mentioned in the blog post into the database in wiki style, the corresponding provenance information Prov_2 using PROVHeart is shown in Fig. 17.

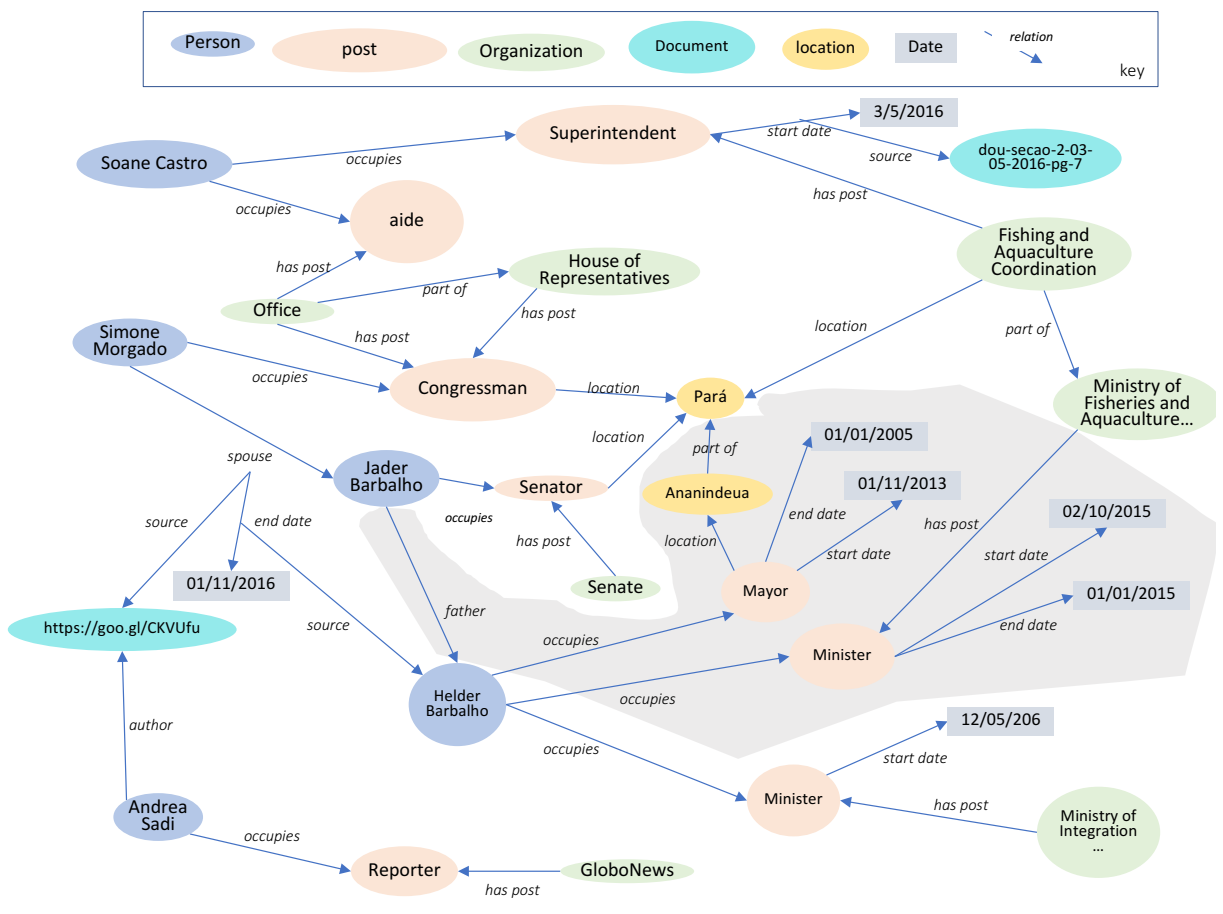

Fig. 15 – An instance of POLARE with facts underlying a blog post

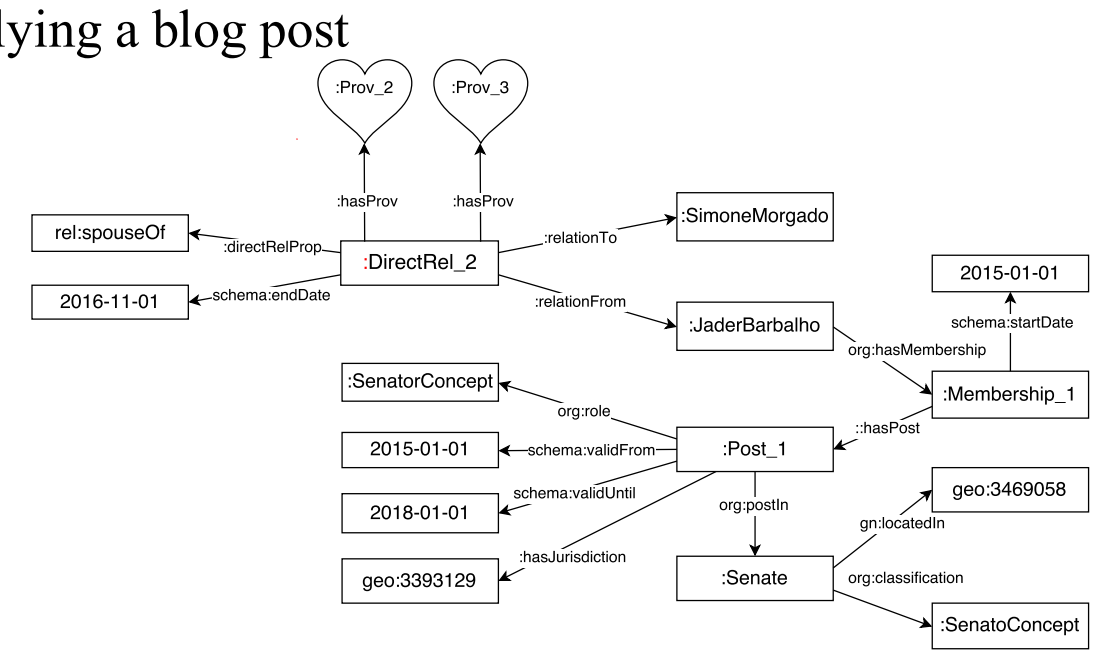

Fig. 16 – Kinship relation modeled using POLARE

In this provenance information, Laufer has chosen to interpret the blog post as stating that the spouse relationship has an end date, in spite of the slight "conflict" in the text. It would be possible to create a separate claim for the end date should this be considered relevant and important.

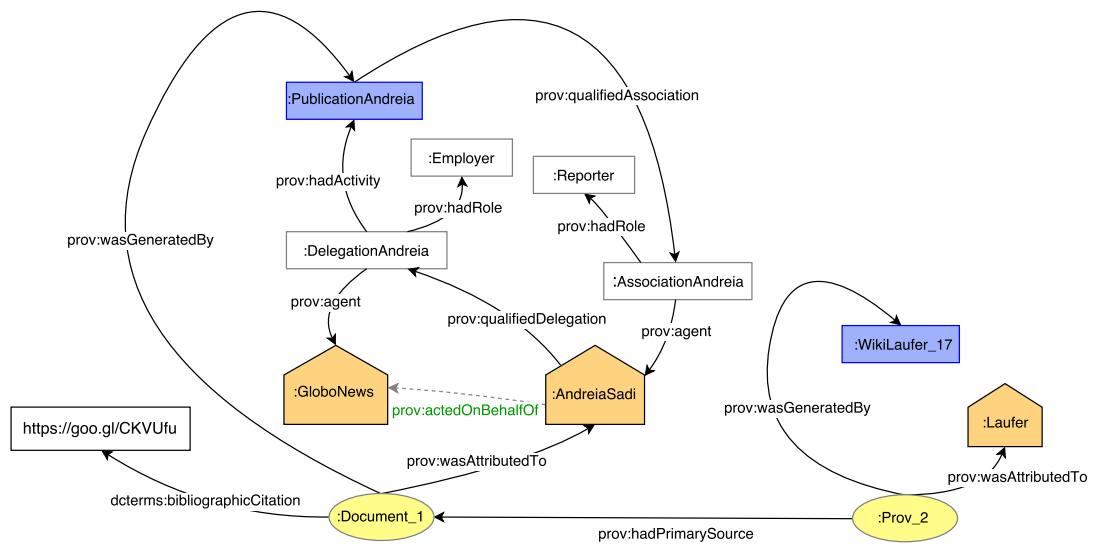

Fig. 17 - First provenance information about blog post (Prov_2).

Suppose now that another agent, Daniel, decided to check whether this kinship information is accurate. Short of trying to find a court document stating that they are officially divorced, Daniel searches for additional sources for this claim, and finds a news story stating the same fact, and enters it into the database. This becomes a second provenance information for the claim in the database, which is represented in Fig. 18Fig. 16 as the Prov_3 PROVHeart instance.

As mentioned, the actual trust process is outside the scope of this paper, but nevertheless how the provenance information can be used in a trust process can be outlined. In the trust process, the consumer of this data must apply her/his own policies in order to accept the truth of the statement.

In the example, one policy could be to simply accept it (meaning, to use it in her/his computations or decisions) based on her/his trust on the recording agent, e.g., Laufer. A second policy, a bit more cautious, would be to accept it based on the agent responsible for the primary source, Andréia Sadi. At this point, it is not possible to continue the chain of provenance, because there is no additional provenance information given in the blog post. A different kind of policy can be used, however. Analogously to journalistic principles of obtaining multiple independent sources about the truth of some claim, the consumer could decide to trust this claim based on the fact that there are two different provenances given, although, strictly speaking, s/he should also check whether Agents Andréia Sadi and Murilo Ramos are independent.

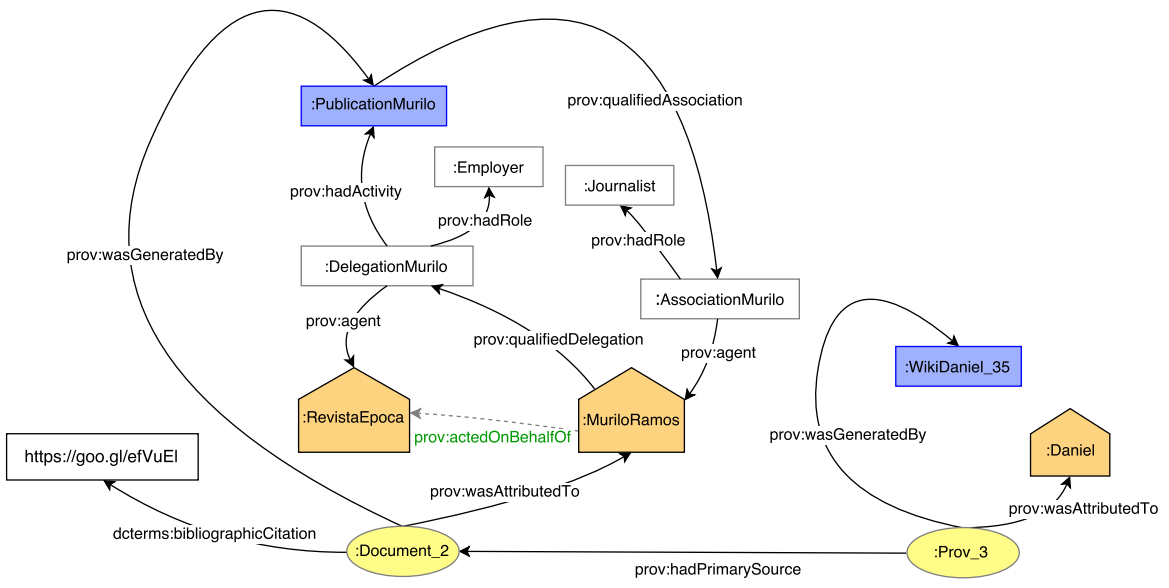

Fig. 18 - Second provenance information about blog post (Prov_3).

## 4. Implementation approach

Rather than present implementation details, this section discusses the strategy followed to populate the database.

### *4.1. KG population*

Given the domain-specific nature of the SLNP KG, a "focused crawling" strategy was adopted, complemented with a user-friendly interface to allow entering individual information items.

The focused crawling approach starts with data dumps provided by official sources, such as the House of Representatives, the Senate, the Supreme Court, etc. These databases provide an initial set of Agents (Persons and Organizations), and instances of some the various types of relations described in Section 2.2. Starting from each Agent in the KG, other sources are searched for additional information, e.g., IRS registry for Organizations, electoral campaign contribution and expenses provided by the Electoral Court, etc… These sources provide instances of relations in which this Agent participates, so the other involved Agents are added to the database, and the process is repeated.

### *4.2. URL Minting*

A key aspect in building the KG is the minting of URIs to uniquely identify each of the resources in it. Most sources used do not provide a standard URI to refer to the entities of interest, and don't provide or use a property that could be used as a global identifier – they typically provide and internal identification key that has no meaning outside their own database.

Given our goal of building an integrated KG, it was decided to generate a random unique identifier for each resource. Whereas URIs using such identifiers are opaque (to human beings), they are a straightforward way to ensure unique keys.

### *4.3. Identity management*

Given the way the POLARE ontologies are defined, there are many types of resources that created to represent information. The actual URLs of most are meaningful only within the SLNP KG, e.g., the URL of a *Membership* class instance standing for a reified relation. On the other hand, instance of Agent – Persons and Organizations – are prima-facie candidates to be referred to by external sources.

A common challenge faced when integrating several data sources into a KG is the disambiguation or identity management of resources, also called entity resolution. There are numerous approaches to address this automatically (see [12] for a survey).

Given the controlled way in which the SLNP is built and maintained, a different strategy is used. For each type of Agent, an "Identity Table" is maintained. Each entry in this table corresponds to an entity, and contains a list of groups of properties and their values that were used in some ingestion process to identify that entity. For example, ("CongressmanName", "Date of Birth", "Social Security Number", etc... In addition, some additional properties are stored which may serve as a "partial key" – in the sense that most entities have a unique value, but it is still possible that some duplication occurs – e.g., "Full Name". It should be remembered that the sources being used for bulk ingestion are, themselves, free of duplicate identities.

When adding a new source, a manual comparison is made with the source's available properties, to identify if the identity table has an equivalent property. If such a property exists, the values are compared using various similarity functions. If more than one match is found, the resolution is done offline by looking at additional available information to try to resolve it.

If no equivalent property exists, a partial key is used, and multiple matches are handled offline, manually. Additional properties that function as keys in the imported database are added to the table. For each group of properties, provenance information of the source database where they function as keys is maintained.

As a result, this identity table can be used as a "owl:sameAs" database between different databases, keeping in mind that they are most often not linked data.

For the sources integrated so far this approach has been successful, requiring manual intervention in the order of only dozens of cases.

When entering information piecewise using the crowdsource interface, the program itself guides users in such a way that duplication is greatly minimized. When entering new information, users are required to identify existing instances by either making queries or browsing before creating new ones, and only when they cannot find them are they allowed to create new instances. Before committing a new value to the KG, a second similarity check is made after the property values have been entered by the user. If duplicates are found, users are required to provide property values that will differentiate them, or else the new entry is discarded and the identity of the found entity used.

## 5. Conclusions

### *5.1. Related Work*

Two projects have proposed similar ontologies as POLARE - the Poderopedia Project[14] and the Popolo Project[15] Poderopedia has built a database about Political Agents allowing instantiations for different countries. It is a Linked Data database using the PoderVocabulary[16], which shares similar concepts with POLARE regarding Persons and Organizations, in some cases reaching a finer grain modeling of some relations, notably between Organizations. It took the approach of using OWL as much as possible, so several of the concepts modeled using SKOS in Polare are modeled as OWL classes. The PoderVocabulary does not detail legal actions, elections nor transactions, and does not have the Referral relation.

The Popolo Project is an initiative to define data interchange formats and data models for governments, in the context of Open Government. They define a set of classes that cover, basically, the specification of Persons and Organizations, using Posts and Memberships to relate Persons to Organizations. Popolo also defines a set of classes related to voting processes. The main principle of Popolo is reuse. The classes are defined as a set of properties of well-known vocabularies, including ORG, FOAF, geonames, schema.org, etc. POLARE defines the linking between Persons to Organizations in a similar way. POLARE introduces a set of relations that are not contemplated by Popolo, as family relations, referrals, legal actions or the elections. Popolo does not include provenance information.

The LittleSis project[17] has created a database with similar goals as the SLNP KG. They also use a combination of automated ingestion processes and human-curated and verified data. The major differences are the fewer types of relations represented (e.g. no legislative activity, legal actions), the lack of explicit queriable provenance data included, and an editorial content control by the project staff. In the SLNP KG, the provenance (and other) information is intended to be employed by users to apply their own trust policies in determining what they consider reliable information, as opposed to relying on the policies of the editors of the database maintainers, as is done for LittleSis.

### *5.2. Future Work*

Work on the SLNP KG is continuing along several directions.

First, existing work on general NLP is being specialized to extract the specific relations used in the KG. One notable example is extracting family relations and membership relations with organizations from published CVs. This will be combined with similar information extracted from social networks.

Second, improved methods for (semi)automated entity resolution are being investigated. One promising approach is examining the use of automated facial recognition, exploiting the fact that persons that are Political Agents have a large number of their photos available on the web.

A third direction being studied is the determination of effective ways to communicate information supporting the trust process in human interfaces.

The interaction with journalists and media people has indicated that for some users, it would be highly desirable to have access control applied to portions of the database. In this scenario, journalists would be willing to enter information into the database that is a result of their work, but make it generally available only after a quarantine period.

One of the main goals of the SLNP project is to enable studies of influence networks and how they operate in a society. The semantic nature of the representation of relations used in the project will be used to allow more domain-specific measures to be defined and computed over the existing graphs, in additional to the traditional topology-based measures defined for networks.

[14] http://www.poderopedia.org
[15] http://www.popoloproject.com/
[16] http://dev.poderopedia.com/documentation/index
[17] https://littlesis.org/

## 6. References

[1] Allcot, H.; Gentskow, M.; Social Media and Fake News in the 2016 Election. Journal of Economic Perspectives 31(2),Spring 2017, pp 211–236

[2] Almendra, V. D. S., & Schwabe, D. (2006). Trust policies for semantic web repositories. In Proceedings of 2nd International Semantic Web Policy Workshop (SWPW'06), at the 5th International Semantic Web Conference (ISWC 2006) (pp. 17-31).

[3] Ananny, M., The partnership press: Lessons for platform-publisher collaborations as Facebook and news outlets team to fight misinformation, Columbia Journalism Review, https://www.cjr.org/tow_center_reports/partnership-press-facebook-news-outlets-team-fight-misinformation.php

[4] Angella J. ;Kim, K; K.P. Johnson, Power of consumers using social media: Examining the influences of brand-related user-generated content on Facebook, Computers in Human Behavior, Volume 58, 2016, Pages 98-108, ISSN 0747-5632, http://dx.doi.org/10.1016/j.chb.2015.12.047. (http://www.sciencedirect.com/science/article/pii/S0747563215303186)

[5] Artz, D; Gil, Y.; A survey of trust in computer science and the Semantic Web, Web Semantics: Science, Services and Agents on the World Wide Web, Volume 5, Issue 2, 2007, Pages 58-71, ISSN 1570-8268, http://dx.doi.org/10.1016/j.websem.2007.03.002.(http://www.sciencedirect.com/science/article/pii/S1570826807000133)

[6] Bizer, C. ; Cyganiak, R.. Quality-Driven Information Filtering Using the WIQA Policy Framework. Web Semantics: Science, Services and Agents on the World Wide Web. 7. 1-10. 10.1016/j.websem.2008.02.005. (2009) - http://www.websemanticsjournal.org/index.php/ps/article/view/157/155

[7] Carlson, A., Betteridge, J., Wang, R.C., Hruschka Jr., E.R., Mitchell, T.M.: Coupled semi-supervised learning for information extraction. In: Proceedings of the Third ACM International Conference on Web Search and Data Mining, pp. 101–110 (2010)

[8] Castelfranchi, C., Falcone, R. Social Trust: A Cognitive Approach. In: Castelfranchi, C.; Yao-Hua Tan (Eds.): Trust and Deception in Virtual Societies. Springer-Verlag (2001).

[9] Castells, Manuel (2010). The Rise of the network society (2 ed.). ISBN 9781405196864. Retrieved 1 December 2016.

[10] Ciampaglia, G. L.; Shiralkar, P.; Rocha, L. M.; Bollen, J.; Menczer, F.; Flammini, A.; "Computational Fact Checking from Knowledge Networks", PLoS ONE, 10(6): e0128193. 2015. http://doi.org/10.1371/journal.pone.0128193

[11] Conroy, N. J., Rubin, V. L. and Chen, Y. (2015), Automatic deception detection: Methods for finding fake news. Proc. Assoc. Info. Sci. Tech., 52: 1–4. doi:10.1002/pra2.2015.145052010082

[12] Christophides, V., Efthymiou, V., & Stefanidis, K. (2015). Entity resolution in the web of data. *Synthesis Lectures on the Semantic Web*, *5*(3), 1-122.

[13] Dutton, W. H. (2009). The Fifth Estate Emerging through the Network of Networks. Prometheus, 27(1), 1–15. http://doi.org/10.1080/08109020802657453

[14] Gerck, E.; Toward Real-World Models of Trust: Reliance on Received Information. Report MCWG-Jan22-1998, DOI: 10.13140/RG.2.1.2913.6727. http://www.safevote.com/papers/trustdef.htm.

[15] Gil, Y.; Artz, D.; Towards content trust of web resources. Web Semant. 5, 4 (December 2007), 227-239. DOI:http://dx.doi.org/10.1016/j.websem.2007.09.005

[16] Grandison, T., & Sloman, M. (2000). A survey of trust in internet applications. IEEE Communications Surveys & Tutorials, 3(4), 2-16.

[17] Groth, Paul, Andrew Gibson, and Jan Velterop. "The anatomy of a nanopublication." Information Services & Use 30.1-2 (2010): 51-56.

[18] Hitzler, P., et al, OWL 2 Web Ontology Language Primer (Second Edition) - https://www.w3.org/TR/owl2-primer/

[19] Holzner B, Holzner L (2006) Transparency in global change: the vanguard of the open society. University of Pittsburgh Press, Pittsburgh, ISBN 9780822972877

[20] Laufer, C; Schwabe, D.; On Modeling Political Systems to Support the Trust Process, Proceedings of the Privacy Online 2017 Workshop (PrivON 2017), co-located with the 16th International Semantic Web Conference (ISWC 2017) Vienna, Austria, Out. 2017. CEUR-WS.org, 2017. v.1951. p.1 – 16

[21] Laufer, C.; Schwabe, D.; Busson, A.; Ontologies for Representing Relations among Political Agents, https://arxiv.org/abs/1804.06015v1

[22] Lehmann, J., Isele, R., Jakob, M., Jentzsch, A., Kontokostas, D., Mendes, P.N., Hellmann, S., Morsey, M., van Kleef, P., Auer, S., Bizer, C.: DBpedia-A large-scale, multilingual knowledge base extracted from Wikipedia. Semant. Web J. **6**(2), 167–195 (2013)

[23] Lenat, D.B.: CYC: a large-scale investment in knowledge infrastructure. Commun. ACM 38(11), 33–38 (1995)

[24] Marwick, A., Lewis, R.. "Media Manipulation and Disinformation Online." Data & Society Research Institute, New York, 2017 - https://datasociety.net/pubs/oh/DataAndSociety_MediaManipulationAndDisinformationOnline.pdf

[25] McKnight, D. H., & Chervany, N. L. (2001). Trust and distrust definitions: One bite at a time. In Trust in Cyber-societies (pp. 27-54). Springer Berlin Heidelberg.

[26] Michael Schudson, The Rise of the Right to Know: Politics and the Culture of Transparency, 1945-1973. Cambridge, MA: Harvard University Press, 2015.

[27] Miles, A., Bechhofer, S. - SKOS simple knowledge organization system reference, World Wide Web Consortium, 18 August 2009 - http://www.w3.org/TR/skos-reference/

[28] Miller, G. A. 1995. WordNet: A lexical database for English. COMMUN ACM 38(11):39–41

[29] Moreau, L., Batlajery, B., Huynh, T. D., Michaelides, D., & Packer, H. (2017). A Templating System to Generate Provenance. Accepted for publication, *IEEE Transactions on Software Engineering*. http://ieeexplore.ieee.org/stamp/stamp.jsp?arnumber=7909036

[30] Nguyen, V., Bodenreider, O., & Sheth, A. (2014, April). Don't like RDF reification?: making statements about statements using singleton property. In Proceedings of the 23rd international conference on World wide web (pp. 759-770). ACM.

[31] Pinyol, I.; Sabater-Mir, J.; Computational trust and reputation models for open multi-agent systems: a review. Artificial Intelligence Review 40:1–25 DOI 10.1007/s10462-011-9277-z (2013)

[32] Raymond, Y.; Sandler, M.; Evaluation of the Music Ontology Framework, Proceedings of ESWC 2012, Lecture Notes in Computer Science, volume 7295, Springer, Heidelberg, 2012.

[33] Ruchansky, N., Seo, S., & Liu, Y. (2017). CSI: A Hybrid Deep Model for Fake News. arXiv preprint arXiv:1703.06959.

[34] Sahlins, M. D. 1960. The origin of society. Scientific American 203(3): 76–87.

[35] Sherchan, W., Nepal, S., and Paris, C. 2013. A Survey of trust in social networks. ACM Comput. Surv. 45, 4, Article 47 (August 2013), 33 pages. DOI: http://dx.doi.org/10.1145/2501654.2501661

[36] Shi, Longxiang, Shijian Li, Xiaoran Yang, Jiaheng Qi, Gang Pan, and Binbin Zhou. "Semantic health knowledge graph:

Semantic integration of heterogeneous medical knowledge and services." *BioMed Research International* 2017 (2017).

[37] Speer, R., Havasi, C.; "Representing General Relational Knowledge in ConceptNet 5, Proceedings of the Eight International Conference on Language Resources and Evaluation (LREC'12), ELRA, 2012.

[38] Suchanek, F.M., Kasneci, G., Weikum, G.: YAGO: a core of semantic knowledge unifying WordNet and Wikipedia. In: 16th International Conference on World Wide Web, pp. 697–706 (2007).

[39] Vrandečić, D., Krötzsch, M.: Wikidata: a free collaborative knowledge base. Commun. ACM **57**(10), 78–85 (2014)

[40] Wardle, C., "Fake News. It's Complicated", FirstDraft News, https://firstdraftnews.com/fake-news-complicated/, accessed on May 12, 2017.